\documentclass[11pt]{article} % use larger type; default would be 10pt

\usepackage[utf8]{inputenc} % set input encoding (not needed with XeLaTeX)

\usepackage{geometry} % to change the page dimensions
\usepackage{graphicx} % support the \includegraphics command and options

\usepackage{ bm }
\usepackage{ textcomp }
\usepackage{ amsmath }
\usepackage{ amssymb }
\usepackage{ authblk }

\usepackage[colorlinks=true, citecolor=blue]{hyperref} % beautiful blue links
\usepackage[numbers,sort&compress]{natbib}
\usepackage{booktabs} % for much better looking tables
\usepackage{array} % for better arrays (eg matrices) in maths
\usepackage{paralist} % very flexible & customisable lists (eg. enumerate/itemize, etc.)
\usepackage{verbatim} % adds environment for commenting out blocks of text & for better verbatim
\usepackage{subfig} % make it possible to include more than one captioned figure/table in a single float
\usepackage{fancyhdr} % This should be set AFTER setting up the page geometry
\usepackage{sectsty}
\allsectionsfont{\mdseries\upshape} % AG

\usepackage[title]{appendix}

\usepackage[nottoc,notlof,notlot]{tocbibind} % Put the bibliography in the ToC
\usepackage[titles,subfigure]{tocloft} % Alter the style of the Table of Contents

\title{Searching for new ferromagnetism precursors in two-dimensional model materials in frame of local force theorem}
\author{I. V. Kashin$^{*}$}
\author{A. Gerasimov}
\author{E. V. Syrnikov}

\affil{\textit{Theoretical Physics and Applied Mathematics Department, Ural Federal University, Mira Str. 19, 620002 Ekaterinburg, Russia}}
\affil{$^{*}$Corresponding author: i.v.kashin@urfu.ru}

\begin{document}
\maketitle

\newcommand{\frc}[2]{\raisebox{2pt}{$#1$}\big/\raisebox{-3pt}{$#2$}}

\newcommand{\MagnTransTemp}{{\mathbb{T}}}

\newcommand{\MagnTransTempNormalized}{\frc{\mathbb{T}}{\mathbb{T}^{*}}}

\newcommand{\regexp}{\mathrm{exp}}

\newcommand{\regsin}{\mathrm{sin}}
\newcommand{\regcos}{\mathrm{cos}}

\newcommand{\energy}{E}

\newcommand{\GkE}{{\cal{G}}^{\sigma}(\energy, \bm{k})}

\newcommand{\GkEup}  {{\cal{G}}^{\uparrow}  (\energy, \bm{k})}
\newcommand{\GkEdown}{{\cal{G}}^{\downarrow}(\energy, \bm{k})}
\newcommand{\GkprimeEdown}{{\cal{G}}^{\downarrow}(\energy, \bm{k'})}

\newcommand{\dE}{dE}

\newcommand{\shortminus}{\text{-}}

\newcommand{\shorteq}{\text{\textdblhyphen \,}}

\newcommand{\sublati}{\tilde{i}}
\newcommand{\sublatj}{\tilde{j}}

\newcommand{\JqZero}{J(\bm{q} \, \shorteq 0)}

\newcommand{\DoS}{\mathrm{DoS}}

\newcommand{\ParagraphTitle}[1]{\textit{#1}}

\newcommand{\n}[1]{\phantom{-} #1}

\noindent
In this work we conduct a numerical search of non-trivial mechanisms, leading to new tendencies towards long-range ferromagnetic ordering in two-dimensional materials. 
For this purpose we employ an original variant of pairwise infinitesimal spin rotations technique to establish the magnetic transition temperature as the rigid function of basic crystal's parameters. 
It favored the numerical optimization of this function using modified genetic algorithm, designed to harvest local extrema. 
It resulted in revealing the moderate metallicity, accompanied by essential orbital anisotropy, as the prime configuration, which provides the most favoring conditions to ferromagnetic ordering, related to double-exchange and superexchange mechanisms.

% --------------------------------------------------------------------

\section{Introduction}

Nowadays the global trend of miniaturization and ecologization in science and industry pushes the progress towards utilizing the low-dimensional physical systems, such as single-layered crystal structures and clusters of individual adatoms. It manifest itself as naturally coming substitution of traditional unit electronic elements~\cite{NewElectronicDevices_1, NewElectronicDevices_2, NewElectronicDevices_3}. Indeed, we observe the tremendous and constantly growing number of theoretical and experimental studies, devoted to discovering novel physical mechanisms and special systems to be applied as the computer memory cell or elementary arithmetic device~\cite{NewCompDevices_1, NewCompDevices_2, NewCompDevices_3, NewCompDevices_4, NewCompDevices_5}.

Speaking about two-dimensional and quasi-two-dimensional materials, one states its remarkable sensitivity to macroscopic external influences. For instance, electric field, being applied to layered structures, was experimentally registered to control magnetic orientation~\cite{ElectricField_1, ElectricField_2, ElectricField_3, ElectricField_4}; electronic and magnetic properties of such systems could be also tuned by strain~\cite{Strain_1, Strain_2}.

Important to note, that magnetic ordering monodriver in 2D materials is generally forbidden by Mermin-Vagner theorem~\cite{Mermin_Wagner_theorem}. Thus any real representative has no chance not to attract a great deal of scientific attention. From theoretical point of view, the long-range magnetic ordering in this case stabilizes owing to composition of several mechanisms. Compensating nature of its net appearance makes possible to lift the restriction. For example, CrI$_3$ monolayer is revealed to be Ising ferromagnet due to effect of Heisenberg isotropic exchange interactions and on-site easy axis anisotropy~\cite{CrI3_Ising_FM_1, CrI3_Ising_FM_2}. 

Along with isotropic interatomic effects, one can but mention the anisotropic ones. Thus, authors of the Ref.~\cite{DMI_2D_1} revealed essential Dzyaloshinskii–Moriya interaction (DMI) in graphene–ferromagnet interfaces due to the Rashba effect. And in transition metal dichalcogenide there was found a possibility to engineering DMI by Fe atoms interaction~\cite{DMI_2D_2}.

Summarizing, we should state that truly comprehensive nature of magnetic ordering in 2D materials makes the finding a new representative to be real challenge. Since fundamental quantum interplays of different microscopic mechanisms could be exhaustively investigated only by means of modern theoretical approaches, we consider promising to employ it for exploration of non-trivial favoring tendencies on the level of electronic structure. This purpose could be manifested as the prime goal of our study.

To meet it we adapt the well-developed theoretical concept, based on local force theorem~\cite{LocalForceTheorem_1, LocalForceTheorem_2}. It enables one to routinely estimate the magnetic transition temperature ($\MagnTransTemp$) of arbitrary crystal, defined on the level of Bravais lattice and tight-binding Hamiltonian~\cite{Goringe_1997}. Having $\MagnTransTemp$ established as the function of crystal's geometry and electron system, we perform the numerical search for extrema. Thus obtained series of model crystals could be treated statistically or individually - to complete the physical interpretation. 

Previous investigations of this regard were focused on general exploration of the interpolated field, formed by known 2D materials~\cite{song2020computational}. Interest was to employ machine learning techniques for seeking the new structures, that pretend to appear stable by means of total energy, obtained from first-principle calculations.

In our work we focus on distinguishing the fundamental trends on the level of the closest interatomic electron interplays, keeping the effective single band interpretation available to be extrapolated on the case of real materials. Due to the point that practical identification of new magnetic driving trends turns to be severely non-straightforward, the highlighting of physical mechanisms and basic characteristics, that appear able to stimulate the ferromagnetic ordering in the reduced crystal field of 2D materials, is assumed to be of practical and theoretical importance.

% --------------------------------------------------------------------

\section{Method}

A numerical scheme, that we design to perform the 2D magnetics search, meaningfully consists of two distinct modules. The first one is to establish the rigid correspondence between crystal's basic parameters and magnetic transition temperature ($\MagnTransTemp$). The second one appears as special modification of popular optimization approach, aimed to autonomously capture and protocolize extrema of thus formulated $\MagnTransTemp$ function. Below we provide the dedicated description of both modules.

% ------------------------------------------------------
\subsection{Estimation of $\MagnTransTemp$}

In order to evaluate $\MagnTransTemp$
we shall start from representation of an arbitrary crystal in frame of tight-binding approximation. The Hamiltonian then contains on-site electron energies and hopping matrices between atom $i$, located in the crystal's unit cell with zero translation vector, and atom $j$ from the unit cell with translation vector $\bm{T}$:
\begin{equation}
    \big[
    H^{\sigma}(\bm{T})
    \big]_{ij} 
    = 
    t_{ij}^{\sigma}
    + 
    \varepsilon_{i}^{\sigma} \, 
    \delta_{ij} \, 
    \delta_{\bm{T} 0} \, ,
\label{TB_Hamiltonian}
\end{equation}
where 
$\big[ H^{\sigma}(\bm{T}) \big]_{ij}$ is $ij$-sector of unit-cell-sized matrix,
$t_{ij}^{\sigma}$ - electron hoppings matrix (we consider only spin-preserving hoppings),
$\varepsilon_{i}^{\sigma}$ - on-site electron energies.

The next step is rewriting this Hamiltonian in reciprocal $\bm{k}$-space:
\begin{equation}
      H^{\sigma}(\bm{k}) = \sum_{\bm{T}} \, H^{\sigma}(\bm{T}) \cdot \regexp(i \bm{k} \bm{T})
      \, .
\end{equation}
It allows one to introduce the Green's functions formalism. For some $\bm{k}$-vector and energy $\energy$ we write
\begin{equation}
      \GkE = \big\{ \energy - H^{\sigma}(\bm{k}) \big\} ^{-1} \, .
\label{GreenEK}
\end{equation}
In order to compose 
on-site $G^{\sigma}_{i}$ and 
inter-site $G^{\sigma}_{ij}$ crystal's Green functions, 
we should perform a summation over the first Brillouin zone:
\begin{equation}
    G^{\sigma}_{i} = 
    \frac{1}{N_{\bm{k}}} 
    \sum_{\bm{k}} 
    \big[ \GkE \big]_{ii}  \, ,
\label{Gi-sigma}
\end{equation}
\begin{equation}
    G^{\sigma}_{ij} = 
    \frac{1}{N_{\bm{k}}} 
    \sum_{\bm{k}} 
    \big[ \GkE \big]_{ij}
    \cdot 
    \regexp(-i \bm{k} \bm{T}_{ij})  \, ,
\label{Gij-sigma}
\end{equation}
where $\bm{T}_{ij}$ is the translation vector between the cells of atoms $i$ and $j$, $N_{\bm{k}}$ is the number of Monkhorst-Pack $\bm{k}$-grid points~\cite{Monkhorst_Pack}, $\energy$ as an argument is omitted for brevity.

To evaluate $\MagnTransTemp$ for arbitrary 2D crystal we consider the mapping of initial electron model into an effective spin one:
\begin{equation}
      {\cal{H}} = 
      -
      \sum_{ij} 
      J_{ij} \,\,
      \bm{e}_i \cdot \bm{e}_j
      \, ,
\end{equation}
where $\bm{e}_a$ is the unit vector of "classic" spin representation,
in the sum each pair of spins is taken twice. 
One can see that if we introduce the small perturbation from ferromagnetic ground state as infinitesimal spin rotations, the second variation of the spin Hamiltonian could be written as:
\begin{equation}
      \delta^2 {\cal{H}} = 
      -
      \sum_{ij} 
      J_{ij} \,
      \Big[
            \bm{\delta^2 e_i} \cdot \bm{e_j}
                  +
            2 \; \bm{\delta e_i} \cdot \bm{\delta e_j}
                  +
            \bm{e_i} \cdot \bm{\delta^2 e_j}
      \Big]
      \, .
\label{SpinEnergySecondVariation}
\end{equation}
Then the local force theorem provides us a possibility to represent this variation at the level of electron Green's functions~\cite{LKAG_1987}. It results in an expression for corresponding pairwise exchange interaction between atoms $i$ and $j$:
\begin{equation}
    J_{ij} = 
    \frac{1}{8 \pi} 
    \int_{-\infty}^{\energy_{F}} 
    \mathrm{Im} \,
    \mathrm{Tr_{L}} 
    \bigg(
    \sum_{\sigma}
    \Delta_{i} G_{ij}^{\sigma} \Delta_{j} G_{ji}^{- \sigma} 
    \bigg)
    \, \dE \, ,
\label{LKAG_Jij}
\end{equation}
where 
$\energy_{F}$ is the Fermi level, 
$\mathrm{Tr_{L}}$ is the trace over orbital indices, 
$\sigma$ is the spin indices ($- \sigma$ denotes the opposite direction),
$\Delta_{i}$ is intraatomic spin-splitting:
\begin{equation}
    \Delta_{i} = 
    \big[ H^{\uparrow}(\bm{T} \shorteq 0) \big] _{ii} 
    - 
    \big[ H^{\downarrow}(\bm{T} \shorteq 0) \big] _{ii} 
    \, .
\label{DeltaEq}
\end{equation}

\begin{figure}[!h]
\centering
\includegraphics[width = 0.49 \columnwidth]{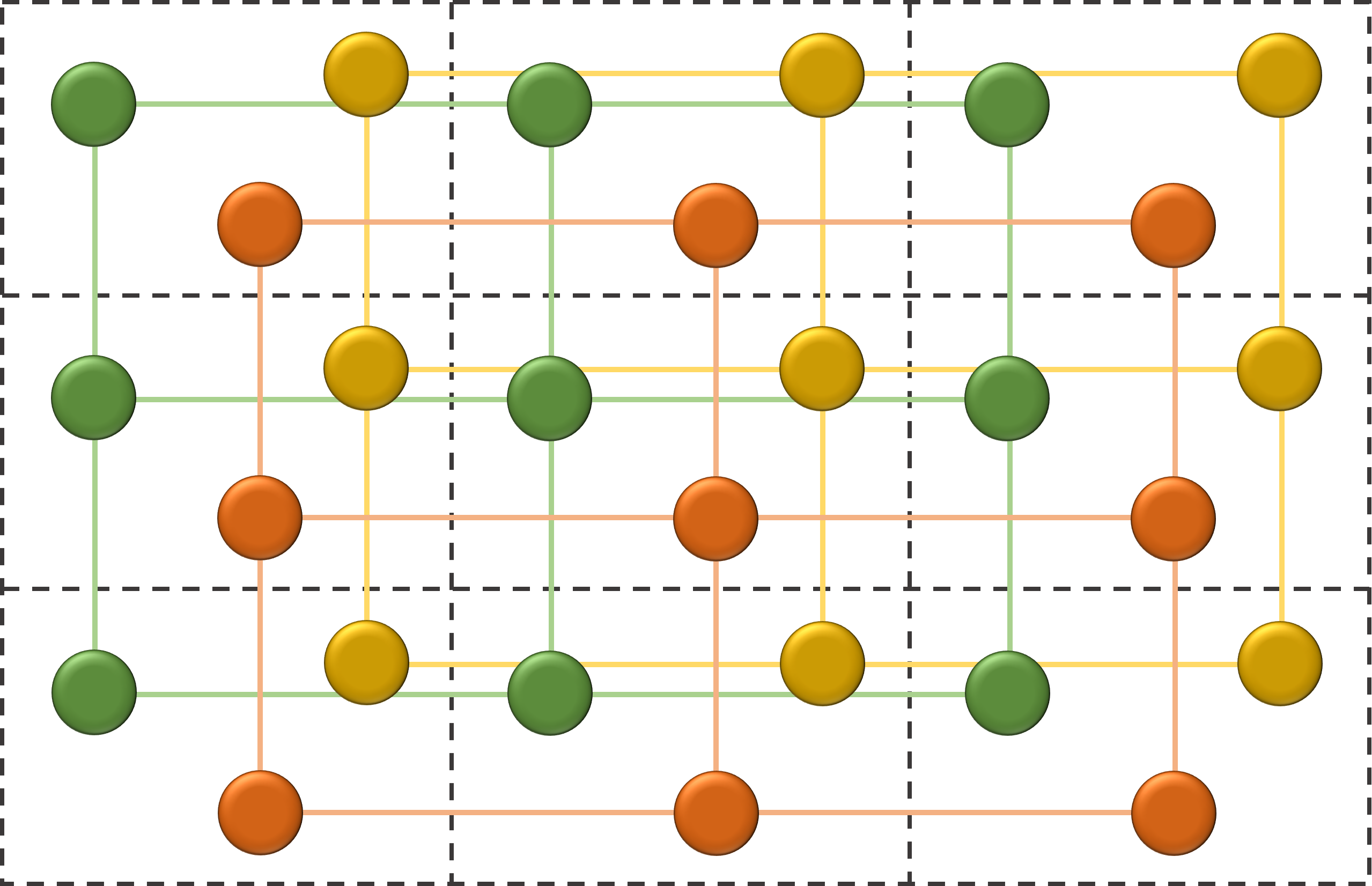}
\caption{Illustration of formal dividing initial crystal structure into sublattices.}
\label{Fig:Sublattices}
\end{figure}

And finally we apply mean-field approximation~\cite{RPA_MFA_MnSi, RPA_MFA_Good_Formulas}, that enables one to find $\MagnTransTemp$ as the largest eigenvalue of the matrix, composed from $J_{ij}$ in a following way. 
Let us formally divide the crystal into sublattices, where each of them contains only atoms with the same position in the unit cell (Figure \ref{Fig:Sublattices}).

Then we define the total exchange interaction between atom $i$ of the $\bm{T} \shorteq 0$ unit cell and all atoms $j$ from the sublattice $p$ as:
\begin{equation}
      S_{ip} = \sum_{j(p) \neq i} J_{ij} 
      \, ,
\label{Sublattice_Sum}
\end{equation}
and express the element of the matrix to be diagonalized~\cite{My_CrO2_PRB}:
\begin{equation}
      L_{ip} = \frac{2}{3} \, S_{ip}
      \, .
\label{T_MFA_Matrix}
\end{equation}

The general problem that rises from our incentive to capture arbitrary crystal, could be manifested as choosing universal approach for estimation of the spatial sum, Eq.~(\ref{Sublattice_Sum}). Indeed, if the material turns to be metallic, owing to strictly delocalized conduction electrons this sum could converge very slowly or even becomes non-convergent. 

To circumvent it we employ the reciprocal space approach, developed in the work~\cite{Kashin_2021_Reciprocal}. The basic idea is to perform a Fourier transform of exchange interactions:
\begin{equation}
    \big[
    J(\bm{q})
    \big]_{ip} = 
    \sum_{\bm{T}_{ij(p)}} J_{ij} \cdot \regexp(i \bm{q} \bm{T}_{ij}) \, .
\label{Jq_Fourier}
\end{equation}
where $\bm{q}$ is the reciprocal space vector, obeying the crystal's symmetry. Then we use the expression, enabling one to calculate $J(\bm{q})$ directly via Green's function:
\begin{equation}
    \big[
    J(\bm{q})
    \big]_{ip} = 
    \frac{N_{\bm{k}}}{8 \pi} 
    \int_{-\infty}^{\energy_{F}} 
    \mathrm{Im} \,
    \mathrm{Tr_{L}} 
    \bigg( 
    \sum_{\sigma} 
    \sum_{\bm{k}} 
    {\cal{A}}^{\sigma}_{ip}  (\bm{k}+\bm{q}) 
    \cdot 
    {\cal{A}}^{-\sigma}_{pi} (\bm{k}) 
    \bigg)
    \, \dE \, .
\label{Jq_Final_Expression}
\end{equation}
where
\begin{equation}
    {\cal{A}}^{\sigma}_{ip} (\bm{k}) = 
    \frac{1}{N_{\bm{k}}}
    \,
    \Delta_{i} 
    \big[ \GkE \big]_{ip} \, ,
\end{equation}
and $p$ denotes the atom from sublattice $p$ in the $\bm{T} \shorteq 0$ unit cell.

We emphasize that for our main purpose it appears sufficient to find only 
$\big[ \JqZero \big]_{ip}$. 
Indeed, one can readily see from Eq.~(\ref{Jq_Fourier}) that it fully contains $S_{ip}$:
\begin{equation}
      S_{ip} = \big[ \JqZero \big]_{ip} - J_{ii} \, \delta_{ip}
\label{S_via_Jq0}
\end{equation}
where 
$\delta_{ip}$ marks inclusion of atom $i$ in the sublattice $p$, and $J_{ii}$ could be found from Eq.~(\ref{LKAG_Jij}).

Therefore, as an approach for $\MagnTransTemp$ estimation we finally state:
\begin{equation}
      \MagnTransTemp =
      \underset{i}{\mathrm{max}} \,
      \{ \lambda_i \}
      \, ,
\label{MagnTransTemp_Expression}
\end{equation}
where $\lambda_i$ is the eigenvalue of $L$, Eq.~(\ref{T_MFA_Matrix}), with $S_{ip}$ taken in the form of Eq.~(\ref{S_via_Jq0}).
% ------------------------------------------------------

% ------------------------------------------------------
\subsection{Optimization}

To accomplish the numerical seeking of $\MagnTransTemp$ extrema at the crystal's parameters space we use the scheme called \textit{Genetic Algorithm (GA)}~\cite{YANG202191}.
Its basic driver could be essentially manifested as mathematical analogue of natural selection process. Here we always have a group of "the best" (by means of particular task) points on the variable parameters space, which produce new "candidates" to challenge them for the place in "the best pool". Thus after a number of generations as "the best" we obtain the most "adapted" points, which have the highest "score".

\begin{figure}[!h]
\centering
\includegraphics[width = 0.8 \columnwidth]{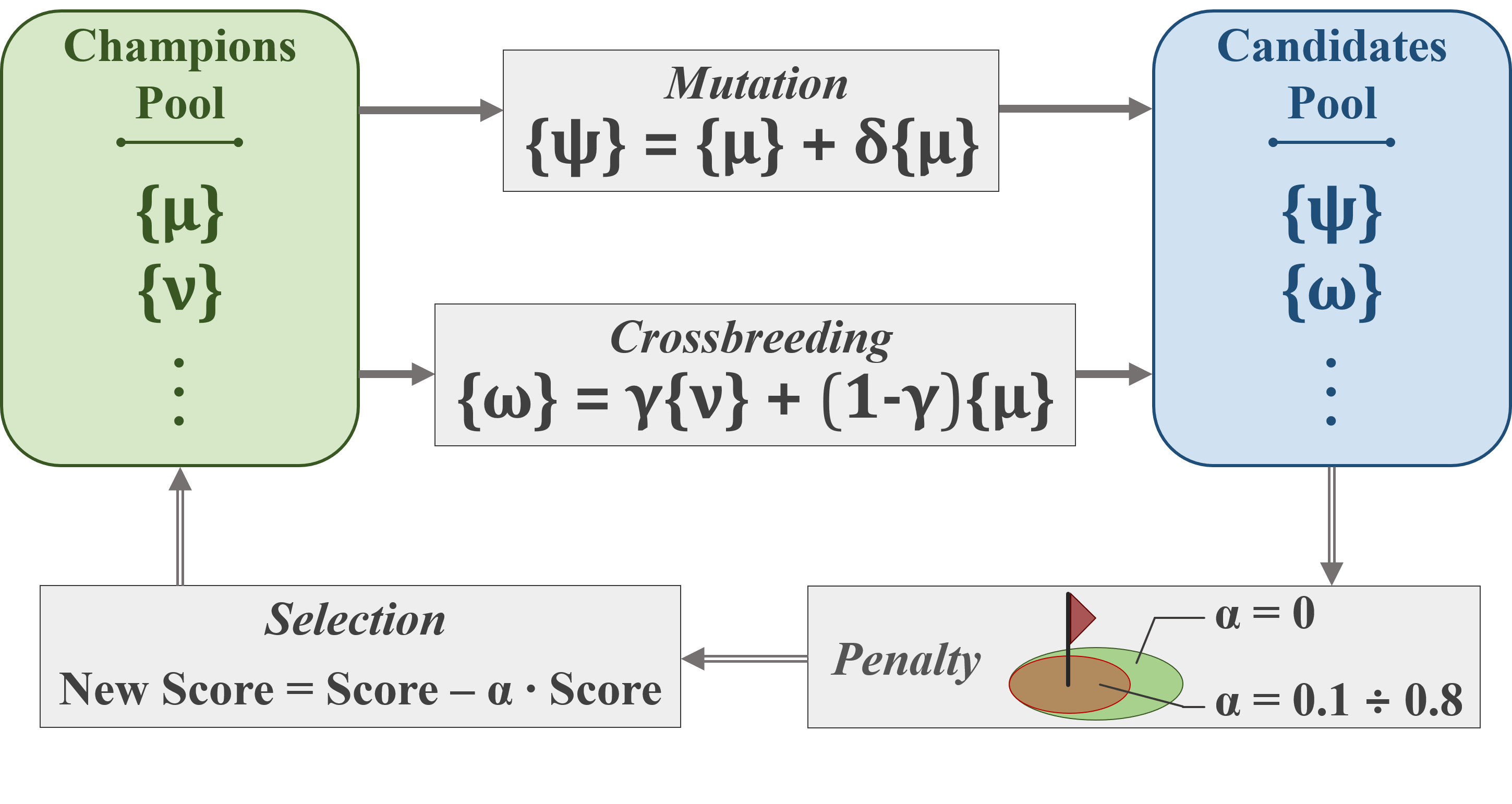}
\caption{Schematic presentation of the modified GA scheme.}
\label{Fig:GA_Scheme}
\end{figure}

For the sake of concreteness in this work we denote "the best" points as "Champions", who form "Champions Pool" and generate "Candidates" to fill the "Candidates Pool" (Figure \ref{Fig:GA_Scheme}).

One of the most popular GA scheme implies two mechanisms of candidates generation. 
The first one is so-called "mutation" and essentially is small variation of randomly chosen champion. This scenario corresponds to small step in the parameters space, providing a possibility to smoothly localize the extrema in the vicinity of the current champion point. 
The second one named "crossbreeding" constitutes the mixture of different champions. If realized as linear superposition of two champions, the resulting candidate appears as the random point on the line, that connects these champions (Figure \ref{Fig:GA_Scheme}).

Elaborating the numerical realization of GA, we should stress the fact that the net diversity of Champions Pool suffers from decaying. Indeed, if we sort the Champions Pool (of the size $N_{ch}$) in descending order by score, introduce the Euclidean distance between two points $\bm{\mu}$ and $\bm{\nu}$ as
\begin{equation}
      F_{\mu, \; \nu} = 
      \sqrt 
       {
        \sum_{g}
        \big( 
             \mu_g - \nu_g 
        \big)^2
       }
       \, ,
\end{equation}
and estimate the total diversity
\begin{equation}
      {\cal{F}} =
            \sum_{\mu = 1}^{N_{ch} - 1} \,
            F_{\mu, \; \mu + 1}
\end{equation}
we can define the convergence criterion of GA as ${\cal{F}} \rightarrow 0$.

In this work we utilize this feature by noting that the convergence reliably signifies the current location of any champion point to be in the closest vicinity to extrema. Therefore, one can mark this location and reset the Champions Pool by random points to proceed searching. In order to boost the numerical performance of such scheme it is necessary to add a mechanism that prevents the further champions to approach already marked extreme points. For this purpose we employ a penalty technique. It implies to severely cut the score of candidates if their distance to mark is less then tunable penalizing radius (Figure~\ref{Fig:GA_Scheme}). If one assumes score to be only positive, the cutting principle could be expressed as:
\begin{equation}
\mathrm{New~Score} = \mathrm{Score} - \alpha \cdot \mathrm{Score} \, ,
\end{equation}
where $\alpha \sim 0.1 \div 0.8$ if penalty is applied and $\alpha = 0$ otherwise.

This modification turns GA into a autonomous machine of extrema listing, for which the finding of all available extrema appears as only a matter of computational time, but not a matter of technical fitting parameters, possessed by optimization algorithm itself.

% ------------------------------------------------------

% ------------------------------------------
\section{Settings of the numerical search}

In this section we formulate the specific task to be accomplished in the work.

\begin{itemize}

\item
Due to the fact a great deal of known 2D magnetics have transition d-metal as the source of magnetism, we set the atom model to contain five orbitals. 

\item
The basis of the Hamiltonian, Eq.~(\ref{TB_Hamiltonian}), is assumed specially adjusted to have no explicit interactions between intra-atomic orbitals. It sets diagonal $5 \times 5$ sectors of $H^{\sigma}(\bm{T} \shorteq 0)$ matrix to be of diagonal form and containing only on-site electron energies.

\item
We consider only case of crystals with the central symmetry:
\begin{equation}
      H^{\sigma}(\bm{T}) = H^{\sigma}(-\bm{T})
      \, .
\end{equation}

\item
Unit cell of the crystal could contain any number of atoms. Atoms are assumed to be equal to each other. 

\item
The hopping matrices are set isotropic and central-symmetric:
\begin{equation}
      t^{\sigma}(\bm{R}_{ij}) = t^{\sigma} \big( |\bm{R}_{ij}| \big)
      \, .
\end{equation}
For the regular crystal structure it means that for each coordination sphere we have one $5 \times 5$ hopping spinor-matrix.

\end{itemize}

According to this frame, arbitrary 2D crystal under our consideration could be defined by the following independent parameters:
\begin{itemize}
      \item Two 2D translation vectors of Bravais lattice
      \item Number of atoms in the unit cell
      \item Positions of atoms in the unit cell
      \item 5 electron energies for the on-site electrons with spin-up and spin-down
      \item Number of the most distant coordination sphere, at which election hoppings assumed non-zero
      \item Spinor-matrices $5 \times 5$ for election hopping at each relevant coordination sphere 
\end{itemize}

In this work we study three basic geometrical 2D structures: \textit{triangular}, \textit{square} and \textit{hexagonal} lattice. The optimization was applied to their electronic structure, considering non-zero hoppings only between the nearest neighbors (the first coordination sphere). Thus for each structure we have 
10 energies $\varepsilon^{\sigma}$ and 
50 hoppings $t^{\sigma}$ 
to be optimized by means of $\MagnTransTemp$. The search ranges were defined as
\begin{align}
\begin{split}
      \varepsilon^{\uparrow} \, &: \,
      [ -2.0    ,   0.0 ]~\mathrm{eV}
            \\
      \varepsilon^{\downarrow} \, &: \,
     \phantom{-} [ 0.0    ,   2.0 ]~\mathrm{eV}
            \\
      t^{\sigma} \, &: \,
      [ -0.5    ,   0.5 ]~\mathrm{eV}
\end{split}
 \,\, .
\end{align}
Due to energies variation we set the Fermi level as fixed zero. Monkhorst-Pack grid is chosen to be $6 \times 6$. 

Regarding GA, we perform the numerical search with 80 Candidates and 50 Champions. As the mutation procedure one arbitrary variable is to be modified by $\pm 0.01$. The threshold value of the total diversity ${\cal{F}}$ was chosen to be 1 eV. Penalizing radius around marked extrema we set as 1 eV and corresponding $\alpha = 0.5$.

We also emphasize that thus defined numerical search scheme captures both positive and negative $\MagnTransTemp$, which indicates stability and instability of FM order, correspondingly. It leads us to practically consider the optimization of $|\MagnTransTemp|$, keeping the sign as an attribute to divide the class of stable FM model materials.
% ------------------------------------------

% ------------------------------------------------------------------------
\section{Results and Discussion}

At first we state that, due to exhausting character of extrema protocolizing, possessed by suggested GA technique, it is to be processing until the total diversity finally fails to cross its threshold value.
In considered case it appeared sufficient to perform $10^5$ generations to reach this saturation.
Thus we obtained distinct 
               4562   ,   3797 and   1699
crystals of triangular, square and hexagonal structure, correspondingly.
Taking into account the model character of such configurations, the real values of $\MagnTransTemp$ commonly appear very large $\sim 10^4$ eV.
Thus for the sake of exemplariness we present sets of data normalized by corresponding maximal $\MagnTransTemp^{*}$.

% ------------------------------------------------------
\subsection{General overview}

% ------------------------------------------------------

\paragraph{\ParagraphTitle{Basic characteristics.}}
As the basic distinguishing driver we consider characteristics, that experimentally or theoretically relate to the magnetic ordering of real materials. 
The central one is the atom's total magnetic moment per orbital:
\begin{equation}
      \tilde{M}_{i} 
            = 
      \tilde{N}^{\uparrow}_{i} - \tilde{N}^{\downarrow}_{i}
            =
      -\frac{1}{5 \pi}
      \int_{-\infty}^{\energy_{F}} 
      \mathrm{Im} \,
      \mathrm{Tr_{L}}
      \big[
            G^{\uparrow}_{i}
            - 
            G^{\downarrow}_{i}
      \big]
      \, \dE 
      \, ,
\end{equation}
where $\tilde{N}^{\uparrow}_{i}$ and $\tilde{N}^{\downarrow}_{i}$ are corresponding total atomic occupations per orbital.

According to Eq.~(\ref{DeltaEq}), the intraatomic spin splitting is assumed to play significant rule in magnetic ordering. Indeed, particularly $\Delta_{i}$ sets the on-site magnetism if introduced on the level of Hartree-Fock numerical approach. Thus, we consider its net value per orbital 
\begin{equation}
      \tilde{\Delta}_{i} =
      \frac{1}{5} \,
      \mathrm{Tr_{L}}
      \big[
            \Delta_{i}
      \big]
      \, .
\end{equation}
Then the crystal's metallicity could be qualitatively described by total density of electron states (DoS) on the Fermi level per orbital:
\begin{equation}
      \tilde{D}^{\sigma}_{i} = 
      \frac{1}{5} \,
      \DoS^{\sigma}_{i}(\energy_{F}) =
      -\frac{1}{5 \pi} \,
      \mathrm{Im} \, 
      \mathrm{Tr_{L}}
      \big[
            G^{\sigma}_{i} (\energy_{F})
      \big]
      \, .
\end{equation}
Along with the value itself, it appears useful to indicate whether we have a dominant DoS peak on the Fermi level or not. For this purpose we denote normalized $\DoS(\energy_{F})$ for an electron with spin $\sigma$ on the orbital $a$ of atom $i$ as:
\begin{equation}
      \big[
            \DoS^{\sigma}_{ai}(\energy_{F})
      \big]^{*} =
            \frac
            {\DoS^{\sigma}_{ai}(\energy_{F})}
            {     \underset{\energy}{\mathrm{max}} \,
                  \big\{
                        \DoS^{\sigma}_{ai}(\energy)
                  \big\}    
            }
            \, ,
\end{equation}
and then take it orbitally averaged:
\begin{equation}
      \tilde{ {\cal{D}} }^{\sigma}_{i} = 
      \frac{1}{5} 
      \sum_{a}
            \big[
                  \DoS^{\sigma}_{ai}(\energy_{F})
            \big]^{*}
            \, .
\end{equation}

The results are composed in Figure 
\ref{Lattices_Basics}.

It can be seen that general landscape are revealed highly non-trivial. Indeed, simple localized type of magnetism, which could be exhaustively described by means of Heisenberg spin model, implies as anticipation the large atomic magnetic moment (about 1~$\mu_B$ per orbital) and $\tilde{\Delta}_{i}$ (about 4~eV per orbital), accompanied by insulating character of electron structure ($\tilde{D}^{\sigma}_{i}~\approx~0$).
From other hand, band-sourced magnetism of Stoner model~\cite{Stoner} manifests itself as opposite pole of expectations:
relatively small localized magnetic moment and intraatomic spin splitting, intense metallic behavior: 
$\tilde{ {\cal{D}} }^{\sigma}_{i}~\approx~1$.

On the contrary, obtained distributions explicitly show a comprehensive nature of the magnetism promoters. 
As an explicit feature one can state the principally metallic character of thus found crystals. However, it does not make a trend towards maximizing the density of electron states on the Fermi level (see distribution of $\tilde{ {\cal{D}} }^{\sigma}_{i}$). 
In this view the moderate values of 
$\tilde{\Delta}_{i}$ (which basically controls the spin-polarized band shift) and 
localized magnetic moment one can find quite expectable. 
Electron occupation of $\tilde{N}^{\uparrow}_{i}~\approx~0.8$ and $\tilde{N}^{\downarrow}_{i}~\approx~0.2$ per orbital 
signifies that distinct magnetic driving mechanisms lie in the field of double-exchange
~\cite{PhysRev.81.440, PhysRev.100.675, PhysRev.118.141} 
and superexchange~\cite{PhysRev.115.2} interplays, as well as interactions of higher order~\cite{PhysRevB.92.144407, PhysRevLett.82.2959}.

% ------------------------------------------------------

% ------------------------------------------------------

\paragraph{\ParagraphTitle{Anisotropic effects.}}

Also extremely important to mention that thus distinguished model crystals show a strong tendency towards non-symmetrical electron hoppings matrices. Despite the general central-symmetry of $t^{\sigma}(\bm{R}_{ij})$, we obtain a remarkable possibility to consider anisotropic atomic couplings as the factor of ferromagnetic ordering. Indeed, such comprehensive orbital interconnections at the level of first-nearest neighbors physics could effectively capture the effects, related to conventional Dzyaloshinskii–Moriya interactions~\cite{DZYALOSHINSKY1958241, PhysRev.120.91, PhysRevLett.76.4825, PhysRevLett.119.167201} and general strong electron-electron correlations~\cite{RevModPhys.78.865}.

\begin{figure}[t]
\centering
\includegraphics[width = 0.75 \columnwidth]{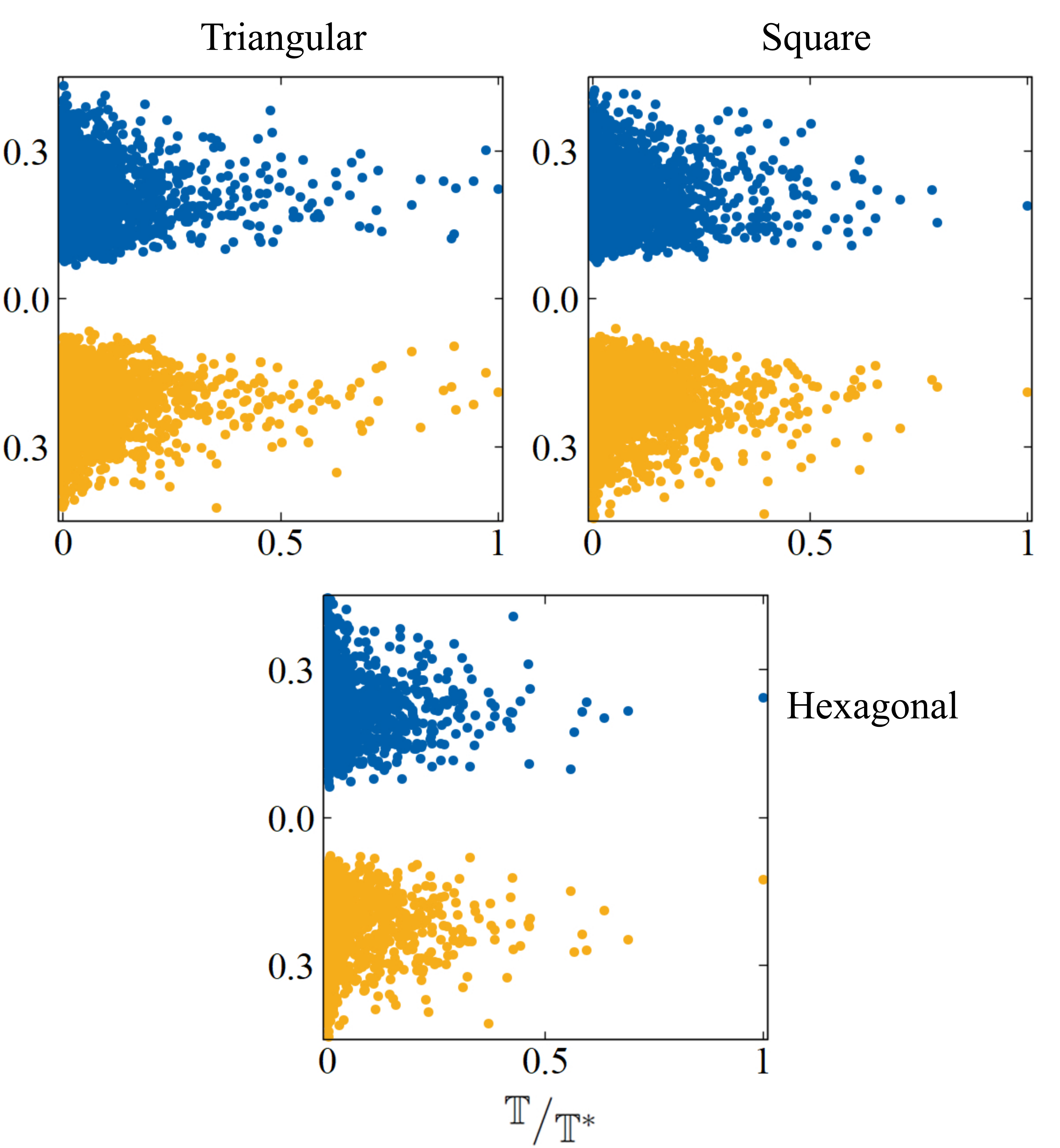}
\caption{Distribution of $\tilde{t}^{\sigma}_{\mathrm{AS}}$ (in eV).}
\label{Fig_Anisotropy}
\end{figure}

To make an illustrative picture we estimate the net asymmetricity per orbital couple as:
\begin{equation}
      \tilde{t}^{\sigma}_{\mathrm{AS}} =
      \frac{1}{25}
      \sum_{ab}
      \big|
            t_{ab}^{\sigma} - t_{ba}^{\sigma}
      \big|
      \, .
\end{equation}
The resulting distributions are presented on Figure~\ref{Fig_Anisotropy}.
As one can see, the mean values demonstrate the moderate behavior. However, in contrast to the basic characteristics, we cannot ascribe to a case of maximal $\tilde{t}^{\sigma}_{\mathrm{AS}}$ a particular physical channel, that alone can favor the ferromagnetic ordering in the material. It signifies that anisotropic effects, which can be tailored, for instance, by strain~\cite{Strain_1, Strain_2}, are reliably expected to increase the robustness of observed FM state, or even to induce the transition into it. 

% ------------------------------------------------------

% ------------------------------------------------------

% ------------------------------------------------------------------------

\subsection{The most stable FM crystals}

\begin{figure}[t]
      \centering
      \includegraphics[width = 0.75 \columnwidth]{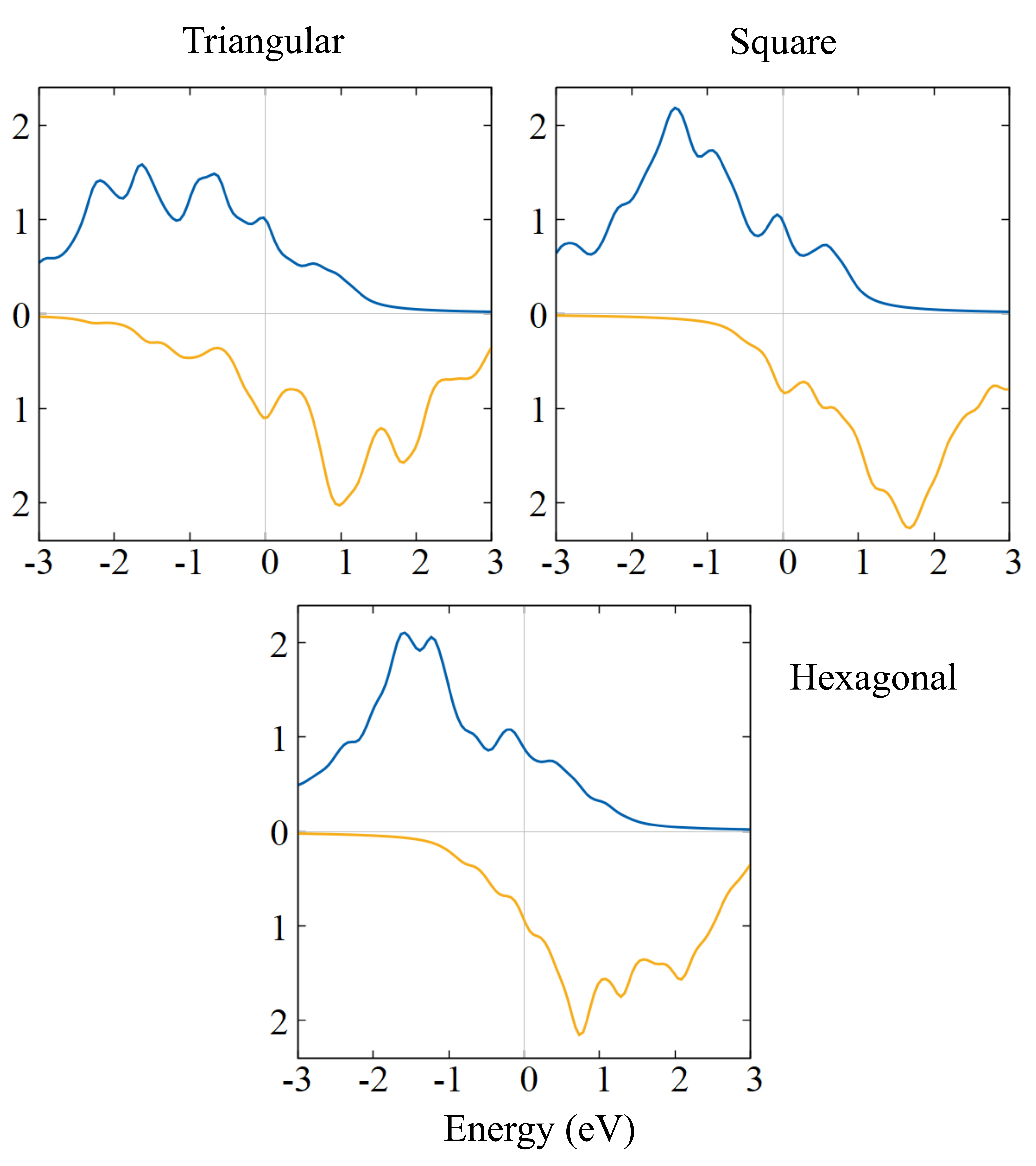}
      \caption
      {
            Total atomic density of states of the most stable crystals found. 
            Blue and orange line denote spin-up and spin-down, correspondingly.
            Fermi level is zero.
      }
      \label{Fig_SuperCrystals_DoS}
\end{figure}

As one can see on Figure~\ref{Lattices_Basics}, for each considered lattice we establish the model crystal, which appears clearly superior by means of magnetic transition temperature
($\MagnTransTemp = \MagnTransTemp^{*}$).
The total atomic density of electronic states (Figure~\ref{Fig_SuperCrystals_DoS}) shows that it indeed inherits the general features of the basic characteristics, possessed by the whole batch.
But extremely important to highlight the tendency of forming a subdominant peak in the closest vicinity of the Fermi level, which serves as the driver of specific moderate metallicity, favoring ferromagnetic ordering by means of double-exchange and superexchange~\cite{PhysRev.100.675, PhysRev.118.141, PhysRev.115.2, PhysRevLett.82.2959, My_CrO2_PRB}.

Table~\ref{SuperCrystals_Table} summarizes the parameters of Hamiltonian, Eq.~(\ref{TB_Hamiltonian}).
We mention that there are both strong anisotropic and spin-polarized effects, which considerably provides the FM stabilizing electronic structure, which is expected to indicate additional magnetism precursors in real 2D materials.

% ------------------------------------------------------------------------
% ------------------------------------------------------------------------

% ------------------------------------------------------------------------

\section{Conclusions}

In this work we apply a novel numerical technique, based on representation of exchange environment in the reciprocal space, to define fundamental trends towards ferromagnetism in 2D materials with standard geometries.
The moderate conducting character and significant spin-polarization of electron hoppings, possessed by obtained model crystals, signifies the principal necessity of pairwise infinitesimal spin rotations to be employed. It is generally validated by the statement, that spin-polarized hoppings do cause intrinsic incorrespondence between single and pairwise spin rotations approaches while estimating the exchange interactions landscape~\cite{Kashin_2021_Reciprocal}.

Subordinate peaks of the density of states near the Fermi level, observed in the model crystals with maximal $\MagnTransTemp$, indicate the double-exchange and superexchange mechanisms to play fundamentally significant role in ferromagnetic ordering. It was shown, that these mechanisms could be additionally stimulated by anisotropic effects.
The results are expected to provide a considerable step towards deep understanding of such comprehensive phenomenon, as thermally stable FM state of two-dimensional systems.

% ------------------------------------------------------------------------

% -------------------------------------------------------------------

\section{Acknowledgements}

The work is supported by the grant program of the President of the Russian Federation MK-2578.2021.1.2.

% -------------------------------------------------------------------

% -----------------------------------------------------------------

\newpage

\begin{appendices}

\section{}

\begin{figure}[!h]
\centering
\includegraphics[width = 0.885 \columnwidth]{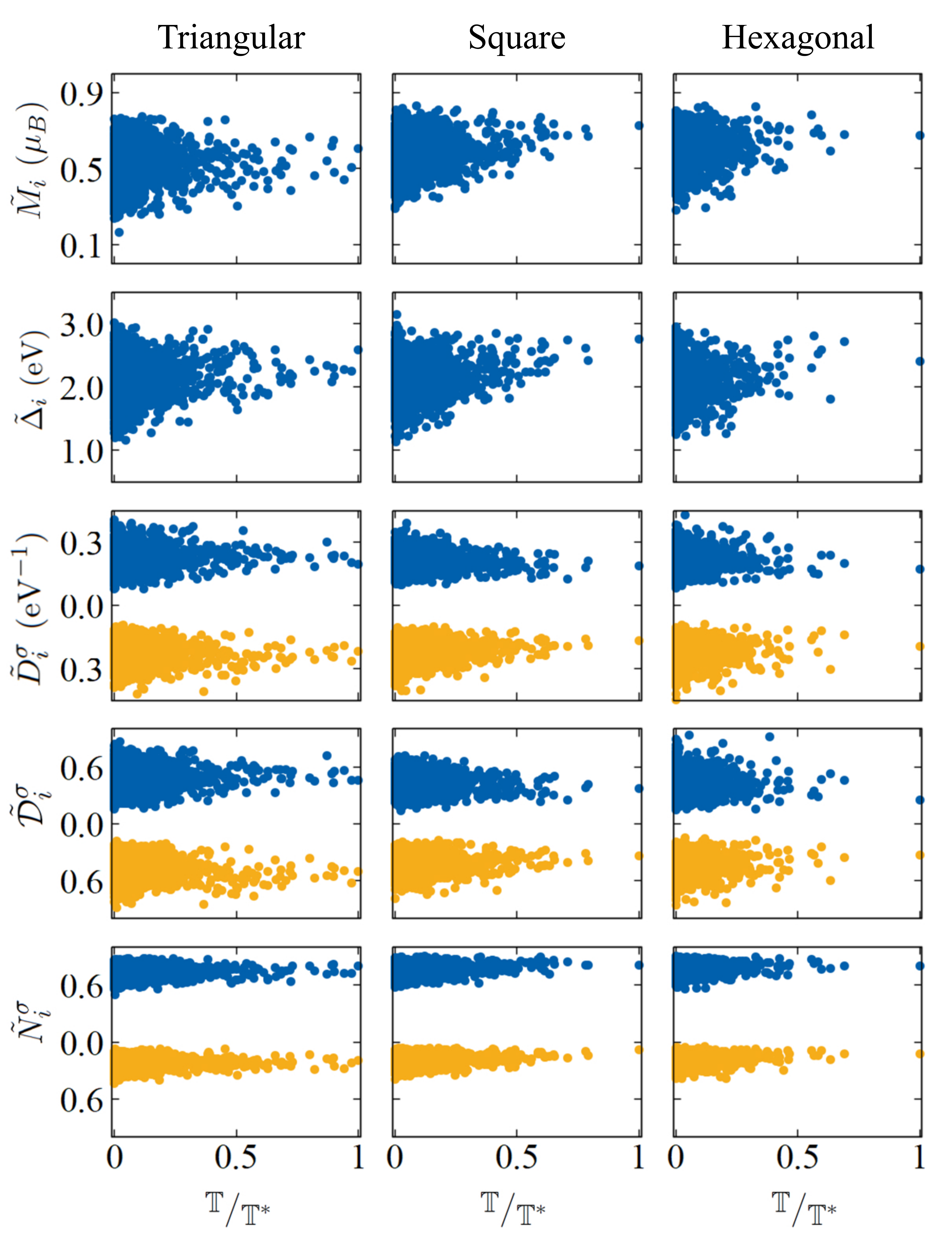}
\caption
{
      The basic characteristics of obtained model crystals.
      For spinors the blue and orange points denote spin-up and spin-down, correspondingly. 
      For \textit{triangular} lattice (\textit{left})   $\MagnTransTemp^{*} = 5.1 \cdot 10^4$ eV.
      For \textit{square}     lattice (\textit{center}) $\MagnTransTemp^{*} = 3.2 \cdot 10^4$ eV.
      For \textit{hexagonal}  lattice (\textit{right})  $\MagnTransTemp^{*} = 2.1 \cdot 10^4$ eV.
}
\label{Lattices_Basics}
\end{figure}

% ====================================
%\begin{eqnarray}
%      \tilde{M}_{i}          ~(\mu_B)      \\
%     \tilde{\Delta}_{i}     ~\mathrm{(eV)}         \\
%     \tilde{D}^{\sigma}_{i} ~\mathrm{(eV^{-1})}      
%      \tilde{ {\cal{D}} }^{\sigma}_{i}     \\
%      \tilde{N}^{\sigma}_{i}             \\
%      \MagnTransTempNormalized
%\end{eqnarray}
% ====================================

\newpage

\begin{table}
\centering
\caption{Hamiltonian parameters of the most stable FM crystals found.}
\renewcommand{\arraystretch}{1.4}
\begin{tabular}{ | c | c | c | c | }
      \hline
            Parameter                       &
            \textit{Triangular lattice}     &
            \textit{Square lattice}         &
            \textit{Hexagonal lattice}      \\
      \hline
           $\varepsilon_{i}^{\uparrow}$~(eV) 
                              &
           $\begin{smallmatrix}
                  -1.5  &  -1.6   &   -1.7   &   -1.5   &    -0.9    \\
            \end{smallmatrix}$
                              &
           $\begin{smallmatrix}
                  -1.5  &  -1.5   &  -0.6    &   -1.0   &    -1.5    \\
            \end{smallmatrix}$
                              &
           $\begin{smallmatrix}
                  -0.7  &  -1.5  &   -1.3    &   -1.7   &    -0.9   \\
            \end{smallmatrix}$
                        \\
      \hline
           $\varepsilon^{\downarrow}$~(eV) 
                              &
           $\begin{smallmatrix}
                \n{1.2}   &    \n{1.3}    &    \n{1.5}    &    \n{0.7}    &    \n{1.0}    \\
            \end{smallmatrix}$
                              &
           $\begin{smallmatrix}
               \n{1.2}    &   \n{1.5}     &   \n{1.7}     &   \n{1.6}     &   \n{1.7}     \\
            \end{smallmatrix}$
                              &
           $\begin{smallmatrix}
               \n{0.5}    & 	\n{1.5}     &   \n{1.3}     &   \n{0.7}     &  	\n{1.9}     \\
            \end{smallmatrix}$
                        \\
      \hline
           $t^{\uparrow}$~(eV) 
                              &
           $\begin{smallmatrix}
                    \phantom{-} & & & & \\
                  -0.1       &       \n{0.2}         &      -0.1         &     -0.3      &     -0.3        \\
                    & & & & \\
                \n{0.4}	     &      -0.3	           &    \n{0.1}	       &   \n{0.0}     &      \n{0.1}    \\
                    & & & & \\
                  -0.3       &	-0.3	           &      -0.4	       &   \n{0.0}     &	\n{0.3}    \\
                    & & & & \\
                  -0.3       &	 \n{0.2}         &    \n{0.1}        &   \n{0.0}     &	\n{0.2}    \\
                    & & & & \\
                \n{0.2}      &    	 \n{0.4}         &	-0.1         &	-0.1       &	-0.2       \\
                    & & & & \\
            \end{smallmatrix}$
                              &
           $\begin{smallmatrix}
                    \phantom{-} & & & & \\
                \n{0.2}     &	  \n{0.0}     & 	  \n{0.3}       &	    \n{0.4}      &  	\n{0.0}       \\
                    & & & & \\
                \n{0.1}     &	    -0.2      &     \n{0.3}       &     \n{0.2}      &        -0.2        \\
                    & & & & \\
                  -0.1      &   \n{0.3}     & 	  \n{0.1}       &       -0.4	     &      \n{0.0}       \\
                    & & & & \\
                \n{0.0}     &     -0.2	  &     \n{0.1}       &     \n{0.0}      &      \n{0.2}       \\
                    & & & & \\
                  -0.3      &	    -0.2      &     \n{0.1}       &     \n{0.2}      &      \n{0.1}       \\
                    & & & & \\
            \end{smallmatrix}$
                              &
           $\begin{smallmatrix}
                    \phantom{-} & & & & \\
      -0.3    &  \n{0.0}    &    -0.1     &  \n{0.0}    &  \n{0.4}      \\
                    & & & & \\
    \n{0.1}   &    -0.1     &    -0.1     &    -0.4     &    -0.3       \\
                    & & & & \\
    \n{0.3}   &    -0.1     &  \n{0.2}    &  \n{0.1}    &  \n{0.1}      \\
                    & & & & \\
    \n{0.4}   &  \n{0.0}    &  \n{0.1}    &  \n{0.3}    &  \n{0.2}      \\
                    & & & & \\
      -0.2    &  \n{0.3}    &    -0.3     &  \n{0.4}    &    -0.2       \\
                    & & & & \\
            \end{smallmatrix}$
                        \\
      \hline
           $t^{\downarrow}$~(eV) 
                              &
           $\begin{smallmatrix}
                    \phantom{-} & & & & \\
                  \n{0.2}  &	     -0.4        &    \n{0.2}       &	\n{0.3}     &	\n{0.2}     \\
                    & & & & \\
                 -0.2	&        \n{0.1}       &	-0.2        &	-0.2        &	\n{0.2}     \\
                    & & & & \\
                 -0.4   &        \n{0.1}       &    \n{0.1}       &	-0.1        &	-0.2        \\
                    & & & & \\
                 -0.1   &           -0.1       &	-0.1	      &     -0.3	      &     -0.1        \\
                    & & & & \\
               \n{0.4}  &	      -0.3       &	-0.1        &	-0.2        &	-0.2        \\
                    & & & & \\
            \end{smallmatrix}$
                              &
           $\begin{smallmatrix}
                    \phantom{-} & & & & \\
                  \n{0.1}   &	   -0.1    &	-0.3    &	   -0.3    &	\n{0.1}     \\
                    & & & & \\
                    -0.2    &	 \n{0.3}   &    \n{0.0}   &	   -0.3    &	\n{0.1}     \\
                    & & & & \\
                    -0.3    &    -0.2    &    \n{0.1}   &      -0.3    &      \n{0.1}     \\
                    & & & & \\
                  \n{0.2}   &  \n{0.1}   &    \n{0.2}   &	   -0.2    &        -0.3      \\
                    & & & & \\
                    -0.1    &  \n{0.0}   &    \n{0.1}   &    \n{0.1}   &	\n{0.0}     \\
                    & & & & \\
            \end{smallmatrix}$
                              &
           $\begin{smallmatrix}
                    \phantom{-} & & & & \\
      -0.3   &       -0.3   &        \n{0.0}   &        \n{0.1}   &      \n{0.2}    \\
                    & & & & \\
      -0.2   &       -0.2   &          -0.2    &          -0.1    &        -0.2     \\
                    & & & & \\
      -0.1   &       -0.1   &        \n{0.2}   &        \n{0.2}   &      \n{0.1}    \\
                    & & & & \\
    \n{0.1}  &       -0.1   &        \n{0.2}   &        \n{0.1}   &      \n{0.0}    \\
                    & & & & \\
    \n{0.0}  &     \n{0.2}  &          -0.1    &        \n{0.4}   &      \n{0.0}    \\
                    & & & & \\
            \end{smallmatrix}$
                        \\
      \hline
\end{tabular}
\label{SuperCrystals_Table}
\end{table}

\end{appendices}

% -----------------------------------------------------------------

\bibliography{Bib/Biblio}

\bibliographystyle{iopart-num}

\end{document}